\documentclass[conference]{IEEEtran}
\IEEEoverridecommandlockouts

\usepackage[utf8]{inputenc}
\usepackage{url}
\usepackage{hyperref}

\usepackage{cite}
\usepackage{float}
\usepackage{amsmath,amssymb,amsfonts}
\usepackage{amsthm}

\usepackage{graphicx}
\usepackage{tikz}

\usepackage{tabu}
\usepackage{array}
\newcolumntype{P}[1]{>{\centering\arraybackslash}p{#1}}
\newcolumntype{M}[1]{>{\centering\arraybackslash}m{#1}}

\usepackage[ruled,vlined,algo2e]{algorithm2e}
\usepackage{algorithm,algpseudocode,float}

\usepackage{caption}
\usepackage{subcaption}
\usepackage{textcomp}
\usepackage{xcolor}
\usepackage[english]{babel}
\usepackage{float}
\usepackage{nomencl}
\usepackage{multicol}
\usepackage{multirow}
\usepackage{makecell}

\def\BibTeX{{\rm B\kern-.05em{\sc i\kern-.025em b}\kern-.08em T\kern-.1667em\lower.7ex\hbox{E}\kern-.125emX}}



\begin{document}

\title{Harnessing Kernel Regression for Stochastic State Estimation in Solar-Integrated Power Grids}

\author{Mohammad Ensaf~\emph{Student Member, IEEE}, Masoud Barati~\emph{Senior Member, IEEE}



\thanks{Mohammad Ensaf and Masoud Barati are with the Department of Electrical and Computer Engineering and Industrial Engineering, Swanson School of Engineering, University of Pittsburgh, PA, USA (emails: {mohammad.ensaf, masoud.barati}@pitt.edu).}
}


%
%

\markboth{ }
\maketitle

\IEEEoverridecommandlockouts
\IEEEpubid{\makebox[\columnwidth]{} \hspace{\columnsep}\makebox[\columnwidth]{ }}
\maketitle
\begin{abstract}
The paper presents a Gaussian/kernel process regression method for real-time state estimation and forecasting of phase angle and angular speed in systems with a high penetration of solar generation units, operating under a sparse measurements regime on both sunny and cloudy days. The method treats unknown terms in the swing equations, such as solar power, as random processes, thereby transforming these equations into stochastic differential equations. The proposed method accurately forecasts and estimates both observed and unobserved operating states, delivering forecasts comparable to those of the standard data-driven Gaussian/kernel process for observed system states. Additionally, the method demonstrates improved accuracy with increased observation frequency and reduced measurement errors in the IEEE 14-bus test system.
\end{abstract}

\vspace{-5pt}
%
\vspace{-5pt}
\IEEEpeerreviewmaketitle
\vspace{-5pt}
\section{Introduction}
\IEEEPARstart{I}{n} modern power systems research, real-time state estimation and forecasting are crucial. These methods use data from sensors and solar models to gauge current system state and predict future trends. This data-driven approach is vital for robust, efficient system operation. State estimation focuses on current metrics like grid voltage, while forecasting uses past data to predict future energy needs and solar yields. The main difference between the two is their temporal focus: estimation is on the present, and forecasting is on the future.

In power system estimation, two key data-driven methods are \textit{frequentist} and \textit{Bayesian} statistics. Frequentist approaches often employ techniques like ordinary least squares, ridge, and lasso regression, as well as advanced models like ARIMA, SVMs, RNNs, and CNNs. Bayesian approaches primarily leverage Gaussian processes. These frameworks produce notably different estimation outcomes in power systems \cite{1,2,3}.

Gaussian Process Regression (GPR)\cite{4} is a quintessential Bayesian non-parametric machine learning technique specifically designed for regression tasks. Functioning as a probabilistic model, GPR discerns the relationship between input attributes, frequently denoted as predictor variables, and their respective outcomes, referred to as response variables, via a Gaussian distribution. The crux of GPR lies in its utilization of a Gaussian process, serving as the prior, to delineate the function mapping inputs to their consequent outputs\cite{5}.

GPR is distinguished by its ability to quantify prediction uncertainties, a trait invaluable for state estimation in power systems, particularly when compared with deep learning approaches. While GPR, a non-parametric and probabilistic model, excels in representing diverse functions and handling sparse training data, it may be less adept with high-dimensional datasets inherent to SCADA systems. Deep learning, reliant on neural networks and flourishing with voluminous data, offers heightened accuracy but lacks intrinsic uncertainty quantification. Furthermore, the unpredictability of SCADA and solar monitoring systems, coupled with factors like communication lags, can sometimes favor the ARIMA model, especially during brief measurement intervals, over GPR in forecasting accuracy \cite{6}.

To the best of our knowledge, there is no prior research to reflect the stochasticity of the solar generation resources in the power swing equation. Our primary concern is to model the real-world random fluctuations in solar power due to cloud cover and estimate the dynamic state of the system. SDEs could model the random fluctuations in solar power input more realistically. These equations would incorporate a stochastic term to account for the variability due to factors like cloud cover. One of the main risks of the model is the complexity of the model. The computational complexity would be higher, but as we used statistical kernel regression, the model can provide more accurate and realistic results\cite{7,8}. 

The proposed model explores the use of GPR in estimating states within power systems, emphasizing the significance of kernel functions and the introduction of stochastic terms to the Ordinary Differential Equations (ODEs), which in this investigation are represented by solar power. While both ODEs and Stochastic Differential Equations (SDEs) are considered, a unique feature of the ODE in this context is the incorporation of a stochastic component: solar power. Kernel functions play a crucial role in determining the covariance among data sets and offer diverse potential in capturing the dynamics of the data. Our key insights include:
\begin{itemize}
\item Examine the influence of solar power, on sunny and cloudy days, on the power-swing equation.
\item A SDE model is used to capture the stochastic behavior of the solar power generation to capture the fluctuation of the cloud in a very narrow time horizon. 
\item A comparative analysis of prediction models against true models based on different traditional and ensemble kernels and performance metrics. 
\end{itemize}

Using RKHSs and optimized GPR, our model predicts observable and latent power states, with randomized cross-validation for data-driven covariance. The paper is structured as follows: Section II explains the problem and math model, Section III presents simulation results, and Section IV concludes.


\section{Problem Statement and General Mathematical Formulation} 

\subsection{Discrete Stochastic Differential Equation Framework}

It is pivotal to delineate the structure of the Stochastic Differential Equation (SDE) which underpins our examination. We utilize the SDE's representation as outlined in \cite{9}:

\vspace{-5pt}

\begin{equation}\label{eq:1}
d X_t = f(X_t) \, dt + \sigma(X_t) \, dW_t
\end{equation}

\subsubsection{Properties of the Wiener Process}

Before delving into the discrete representation, let's recall a few properties of the Wiener process \(W_t\):

\begin{itemize}
    \item \(W_0 = 0\): The process starts at zero.
    \item The increments are independent: For \(0\!\leq\!s\!<\!t\), the increment \(W_t - W_s\) is independent of the history \(\{ W_u, 0 \leq u \leq s \}\).
    \item The increments are normally distributed: \(W_t - W_s \sim \mathcal{N}(0, t-s)\).
    \item The paths are continuous: Though they are nowhere differentiable.
\end{itemize}

\subsubsection{Discretizing the SDE and Convergence}

Given the SDE in equation (\ref{eq:1}), we use the Euler-Maruyama approximation method to express it in discrete form:

\vspace{-5pt}
\begin{equation}\label{eq:2}
X_{n+1} = X_n + f(X_n) \, \Delta t_n + \sigma(X_n) \, \sqrt{\Delta t_n} \, \xi_n
\end{equation}

Where \(\xi_n\) is a random sample from a standard normal distribution.

\subsubsection{Estimation of System Dynamics}

Given the data set \((\boldsymbol{X}, \boldsymbol{Y}) = \{ (X_n, Y_n) \}_{1 \leq n \leq N}\), our goal is to deduce the functions \(f\) and \(\sigma\). The relationship between \(X_n\) and \(Y_n\) is:
\vspace{-5pt}
\begin{equation}\label{eq:3}
Y_n = X_{n+1} - X_n
\end{equation}

\vspace{-5pt}
Building upon our model presumptions, the association between \(X_n\) and \(Y_n\) is given by:
\vspace{-5pt}
\begin{equation}\label{eq:4}
Y_n = f(X_n) \, \Delta t_n + \sigma(X_n) \, \sqrt{\Delta t_n} \, \xi_n
\end{equation}

\subsection{Posteriori Estimation Framework}

To infer the unknown functions \( \!f\! \) and \(\! \sigma\! \), we represent \( \!f\! \) as a vector comprised of elements \(\left\{\hat{f}_n\right\}|_{1 \leq n \leq N}\), while \( \sigma \) is depicted as a vector with components \(\left\{\hat{\sigma}_n\right\}|_{1 \leq n \leq N}\). Upon observing or recording both \( \hat{f} \) and \( \hat{\!\sigma}\! \), we deduce the functions \( \!f\! \) and \( \sigma \) through dual kernel regression methodologies, as discussed in \cite{10}. Initially, the estimations \( \hat{f} \) and \( \hat{\sigma} \) are procured. Although \( \hat{f} = f(X) \) and \( \hat{\sigma} = \sigma(X) \) are conditionally independent given \( X \), \( f \) and \( \sigma \) remain mutually independent.

\subsubsection{Extraction of Function \(f\)} 

Let's assume that we have the observed value, denoted as \( \tilde{\sigma} \). The quadratic form of equation (\ref{eq: 5}) concerning \( \tilde{f} \) suggests that we can derive a closed-form solution for the loss function (\ref{eq:7}). To achieve this, we express \( \tilde{f} \) as the only decision variable:
\vspace{-5pt}
\begin{equation}\label{eq: 5}
\tilde{f}_{opt}(\tilde{\sigma}) = K(X,X) \Lambda (\Lambda K(X,X) \Lambda + \Sigma + \lambda I)^{-1} Y .
\end{equation}

\subsubsection{Extraction of Function \(\sigma\)} 

Incorporating the derived function \texorpdfstring{$\tilde{f}_{opt}(\tilde{\sigma})$}{f_opt} from (\ref{eq: 5}) into equation (\ref{eq: 6}), the decision variable of the loss function is solely dependent on \texorpdfstring{$\tilde{\sigma}$}{sigma}. This relationship can be mathematically represented as:

\vspace{-5pt}
\begin{equation}\label{eq: 6}
\begin{aligned}
\mathcal{L}_{red}\left(\tilde{f}_{opt}(\tilde{\sigma}), \tilde{\sigma}\right) = \left(Y - \Lambda \tilde{f}_{opt}\right)^T(\Sigma + \lambda I)^{-1}\left(Y - \Lambda \tilde{f}_{opt}\right) \\
 + \sum_{n=1}^N \ln \left(\tilde{\sigma}_n^2 \Delta t_n + \lambda\right) + \tilde{f}_{opt}^T K(X, X)^{-1} \tilde{f}_{opt} + \tilde{\sigma}^T G(X, X)^{-1} \tilde{\sigma}. \\
\end{aligned}
\end{equation}

To deduce \( \tilde{\sigma}_{min} \), the optimal solution of (\ref{eq: 6}), we can employ optimization techniques such as the gradient descent method. Considering potential numerical errors, we forecast \( \tilde{\sigma} \) as the mean of the Gaussian vector \( \sigma(X) \), contingent on the constraint \( \sigma(X) = \tilde{\sigma}_{min} + Z \), where \( Z \) represents the perturbation vector. Given that \( Z \) elements, \( Z_i \), follow i.i.d. Gaussian distributions with variance \( \gamma \), the approximation yields:
\vspace{-5pt}
\begin{align}\label{eq:7}
&\tilde{\sigma}_{avg} = \mathbb{E}\left[\sigma(X) \mid \sigma(X) + Z = \tilde{\sigma}_{min}\right] \nonumber \\
&\quad = G(X, X)(G(X, X) + \gamma I)^{-1} \sigma_{min}.
\end{align}

\vspace{-4pt}
Subsequently, \( \sigma \) is represented as:
\vspace{-3pt}
\begin{equation}\label{eq:8}
\sigma_{est}(x) = G(x, X)(G(X, X) + \gamma I)^{-1} \tilde{\sigma}_{min} .
\end{equation}

\subsection{Kernel Functions in Gaussian Processes}

GPs are integral to many machine learning tasks, particularly regression and classification. The kernel or covariance function, central to GPs, gauges the similarity between two input points. Various kernels, each with distinct properties, are available in literature. In this section, we explore notable kernels and their formulas.

\subsubsection{Radial Basis Function (RBF) Kernel}
Also known as the Gaussian kernel, the RBF kernel is a popular choice due to its flexibility. Its formula is given by:
\begin{equation}\label{eq:9}
k_{RBF}(x, x') = \exp\left(-\frac{\|x - x'\|^2}{2l^2}\right)
\end{equation}
where \(l\) is the length scale parameter which controls the smoothness of the function.

\subsubsection{Matérn Kernel}
The Matérn kernel, encompassing the RBF kernel as a subset, is described by:

\vspace{-10pt}
\begin{align} \label{eq:10}
k_{M}(x, x')=\frac{2^{1-\nu}}{\Gamma(\nu)} \left( \frac{\sqrt{2\nu}\|x - x'\|}{l}\right)^\nu K_\nu \left( \frac{\sqrt{2\nu}\|x - x'\|}{l}\right).
\end{align}

Here, \( x \) and \( x' \) are input vectors, \( \nu > 0 \) defines function smoothness, \( l \) is a scale parameter, \( K_\nu \) is the modified Bessel function of order \( \nu \), and \( \Gamma(\nu) \) represents the gamma function.

\subsubsection{Rational Quadratic Kernel}
This kernel can be seen as a scale mixture (infinite sum) of RBF kernels with different characteristic length scales. It is given by:
\begin{equation}\label{eq:11}
k_{RQ}(x, x') = \left(1 + \frac{\|x - x'\|^2}{2\alpha l^2}\right)^{-\alpha}
\end{equation}
where \(\alpha\) is the scale mixture parameter.

\subsubsection{Periodic Kernel (ExpSineSquared)}
The periodic kernel is useful for modeling periodic functions. Its expression is:
\begin{equation}\label{eq:12}
k_{P}(x, x') = \exp\left( -\frac{2\sin^2(\pi \|x - x'\|/p)}{l^2} \right)
\end{equation}
where \(p\) is the periodicity of the function.

\subsubsection{Ensemble of RBF and ExpSineSquared}
Combining kernels can often capture complex structures in the data. An ensemble of RBF and ExpSineSquared is given by:
\begin{equation}\label{eq:13}
k_{E}(x, x') = k_{RBF}(x, x') + k_{P}(x, x')
\end{equation}
This ensemble kernel captures both the general trends (via RBF) and periodic variations (via Periodic) in the data.

\subsection{Evaluation Metrics}
To ensure accurate and reliable predictions from our Gaussian Process models, we utilize three fundamental metrics: Mean Squared Error (MSE), Mean Absolute Error (MAE), and the \(R^2\) score. The MSE, represented by \(\text{MSE} = \frac{1}{n}\sum_{i=1}^{n}(y_i - \hat{y}_i)^2\), quantifies the average squared difference between actual and estimated values, with a lower value indicating a better fit. The MAE, given by \(\text{MAE} = \frac{1}{n}\sum_{i=1}^{n}|y_i - \hat{y}_i|\), computes the average absolute difference and is less influenced by outliers. Lastly, the \(R^2\) score or coefficient of determination, expressed as \(R^2 = 1 - \frac{\text{SS}_{\text{res}}}{\text{SS}_{\text{tot}}}\), measures the variance proportion that's predictable, with a score closer to 1 denoting an excellent fit. Collectively, these metrics offer insights into the model's performance, guiding potential refinements and ensuring prediction dependability.

\subsection{Analysis of Power Transmission Network}

This study delves into the transmission infrastructure of an energy facility, wherein the generators are conceptualized based on traditional generator models. A significant source of variability in this framework arises from the solar power integrated with the power-swing equation. With the presumption that each generator in the system adheres to swing equations and the loads operate under a fixed impedance model, the system's dynamic behavior can be comprehensively articulated.

A Langevin equation is postulated to dictate the dynamics of mechanical solar power, with the understanding that its behavior is not deterministically defined. The network's dynamics, as depicted in Figure 1, encompass three standard generators powered by mechanical solar energy, coupled with a singular load. This is mathematically represented by:
\vspace{-5pt}
\begin{equation}\label{eq:14}
\dot{\theta}_{kt} = \omega_B \left( \omega_{kt} - \omega_s \right) 
\end{equation}
\vspace{-10pt}
\begin{equation}\label{eq:15}
2 H_k \dot{\omega}_{kt} = [-D_k \left( \omega_{kt} - \omega_s \right) + P^s_{rt}]- P_{kt}^e + P_{kt}^m
\end{equation}
\vspace{-15pt}
\begin{align} \label{eq:16}
P_{kt}^e - P_{kt}^d &= \sum_{i=1}^N V_{kt} V_{it} \left( G_{ki} \cos \left( \theta_{kt} - \theta_{it} \right) \right. \nonumber \\
&\quad \left. + B_{ki} \sin \left( \theta_{kt} - \theta_{it} \right) \right), k\in \mathcal{N_G}, r \in \mathcal{N_S}. 
\end{align}

Herein, \( \omega_{kt} \) and \( \theta_{kt} \) denote the rotational speed (in rad/s) and voltage phase angle (in rad) of the \( k^{\text{th}} \) synchronous generator, respectively. The constants \( H_k \) (in sec) and \( D_k \) (in p.u) represent the inertia and damping coefficients of the generator, while \( \omega_B \) (in rad/s) signifies the base frequency for per-unit computations. Moreover, \( \omega_s \) (in rad/s) is the synchronous rotational speed, and \( V_{kt} \) characterizes the voltage magnitude behind the impedance of the \( k^{\text{th}} \) generator in p.u. The terms \( G_{ki} \) and \( B_{ki} \) (both in p.u) represent the Y-bus conductance and susceptance matrices, respectively. Lastly, \( P_{kt}^e \) (in p.u) is the generated power injected into the grid, while \( P_{t}^s \) (in p.u) encapsulates the uncertain solar power generation, and \( P_{kt}^m \) (in p.u) is the input mechanical power of the \( k^{\text{th}} \) generator.

\begin{figure}[ht]
\centering
  \centerline{\includegraphics[width=0.5
  \textwidth]{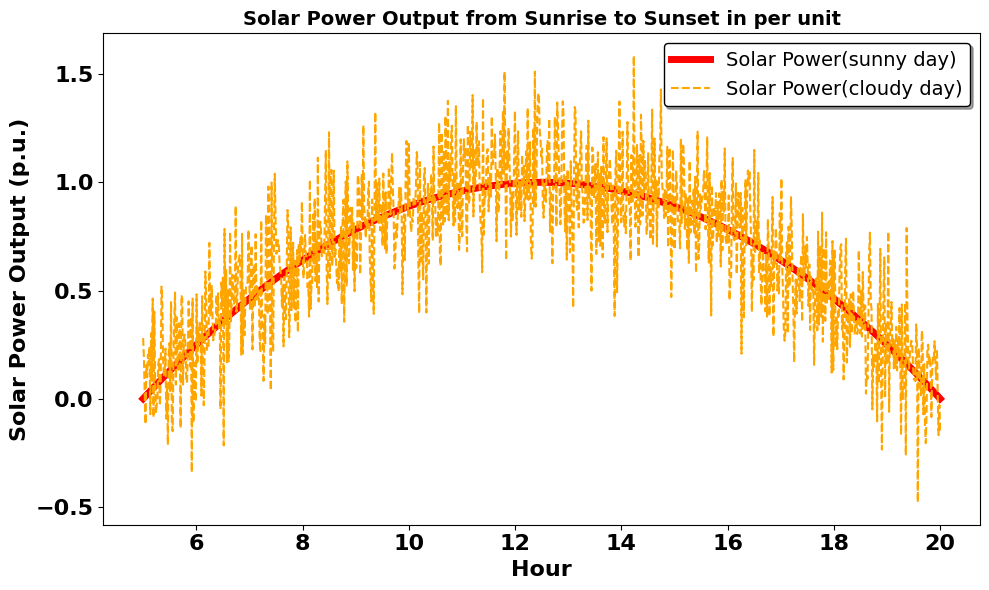}}
  \caption{Solar power output profile from sunrise to sunset: a sunny day vs. a cloudy day.}
  \label{fig: fig1}
\end{figure}

Given 35 successful samples out of the total, this implies an inherent failure in the remaining samples that were not measured, particularly in processes associated with \( \omega_k \) and \( \theta_k \).
 In scenarios devoid of measurements or a definitive deterministic model for solar in sunny day \( P_{r}^s \), predicting the grid's dynamics, given the initial states \( \theta_k(0) = \theta_{k, 0} \) and \( \omega_k(0) = \omega_{k, 0} \) (where \( k \in \mathcal{N_G} \)), becomes challenging. This is further accentuated by \( P_r^s(t) \) representing a time-dependent unknown function, rendering the direct application of equations (\ref{eq:14}), (\ref{eq:15}), and (\ref{eq:16}) impractical without additional solar power data or assumptions. 

\section{Simulation \& Results}

\begin{figure}[ht]
\centering
  \centerline{\includegraphics[width=0.3
  \textwidth]{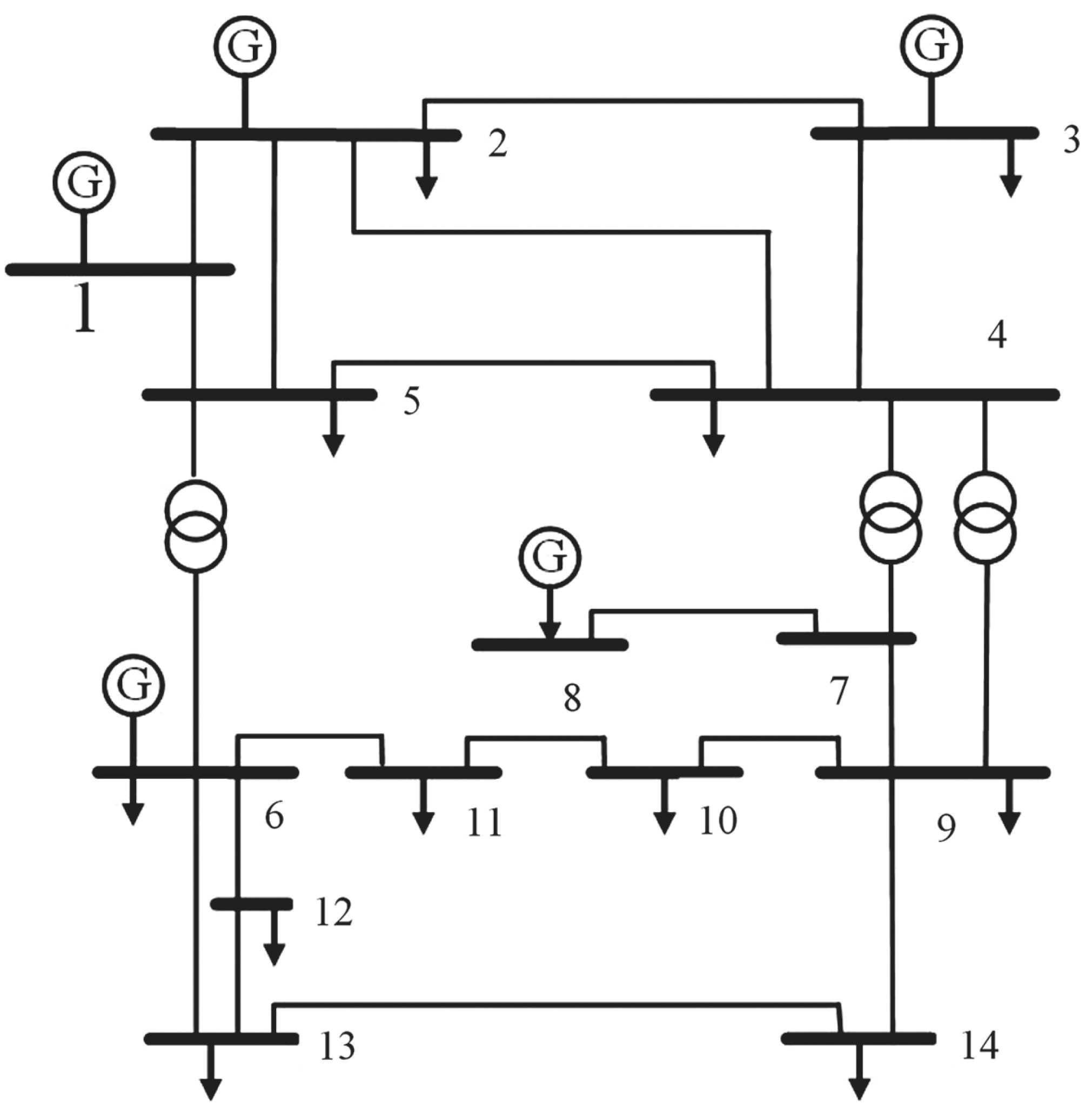}}
  \caption{The IEEE 14 bus system}
  \label{fig: fig2}
\end{figure}

In this study, we focus on the inherent uncertainty introduced by solar power in power systems. To model and study this effect, we employ the IEEE 14-bus system (Comprising 14 buses, it has 5 generators and 11 loads), as illustrated in Fig \ref{fig: fig2}. Given the complexity and interconnectedness of multiple generators in this system, we simplify our analysis by using the equivalent power swing equation \cite{10}. This allows us to study the collective behavior of all generators as if they were a single generator unit. The uncertainty in solar power is captured using a stochastic process for \( P_r^s \), as detailed in (\ref{eq: whitenoise}).
In order to solve the SDE and find accurate solution results, initially, we utilize deterministic ODEs to analyze baseline system behavior under sunny conditions. Subsequently, we incorporated stochastic elements to account for cloud-induced solar fluctuations. 
Fig \ref{fig: fig1} shows two distinct curves depicting the solar power's behavior \( P_{r}^{s} \). The x-axis illustrates the time span from sunrise to sunset, specifically from 5AM to 8PM over a single day. In the study under consideration, two specific solar power scenarios are evaluated: days with clear sunshine and those characterized by cloud cover, as depicted in the provided figure. The objective of this research is to elucidate the impact of varying solar power, based on these two conditions, on the power swing equation as described in equations (\ref{eq: 22}) and (\ref{eq: 23}). In this context, \( P_{r}^{s} \) is introduced to bolster the mechanical power \( P_{k}^m \) in producing the electrical output \( P_{k}^e \).
\vspace{-5pt}

The processes under scrutiny are described as:
\begin{equation}
    P_r^s(t) = \overline{P}_r^s(t) + \delta P_r^s(t)
    \label{eq: whitenoise}
\end{equation}
In this equation, $\overline{P}_r^s(t) > 0$ indicates the average value of the solar power $P_r^s(t)$, whereas $\delta P_r^s(t)$ signifies Gaussian fluctuations centered around zero with a covariance structure explained in equation (\ref{eq: 18}) and (\ref{eq: 19}). Here, $\gamma$ is the reciprocal of twice the variance $\sigma^2$ of the solar power.
\begin{equation}\label{eq: 18}
K(\delta P_r^s(t), \delta P_r^s(t')) = \exp \left(-\gamma ||\delta P_r^s(t) - \delta P_r^s(t')||^2\right) 
\end{equation}
\begin{equation}\label{eq: 19}
K(\delta P_r^s(t), \delta P_l^s(t')) = 0 \text{ for } r \neq l
\end{equation}
For the generation of trajectories, we employ the Euler-Maruyama discretization scheme, which emphasizes the exponential decay in volatility as described by:
\begin{equation}\label{eq: 20}
d \delta P_r^s(t) = \alpha \delta P_r^s(t) dt + \beta \exp \left(-\delta P_r^s(t)^2\right) dW_t
\end{equation}

exponential decay is represented as:
\begin{equation}\label{eq:21}
\begin{gathered}
\delta P_r^s(n+1) - \delta P_r^s(n) = \alpha \delta P_r^s(n) \Delta t \\
\quad + \beta \exp \left(-\delta P_r^s(n)^2\right) \sqrt{\Delta t} \zeta_n.
\end{gathered}
\end{equation}



Equations (\ref{eq: 22}) and (\ref{eq: 23}) are the new version of the Power swing equation that has been incorporated with the solar power. 
\begin{align}
\begin{split}
\frac{d\delta_i}{dt} &= \omega_{B}(\omega_i - \omega_s) 
\label{eq: 22}
\end{split}\\
\begin{split}
\frac{d\omega_i}{dt} &= \frac{\omega_s}{2H_i} \Biggl(P_{mi} - P_{ei} \Biggr) + \frac{\omega_s}{2H_i}\Biggl(-D_k \left( \omega_{kt} - \omega_s \right) + P^{s}_{r}\Biggr)
\label{eq: 23}
\end{split}
\end{align}

\begin{figure}[ht]
\centering
  \centerline{\includegraphics[width=0.5
  \textwidth]{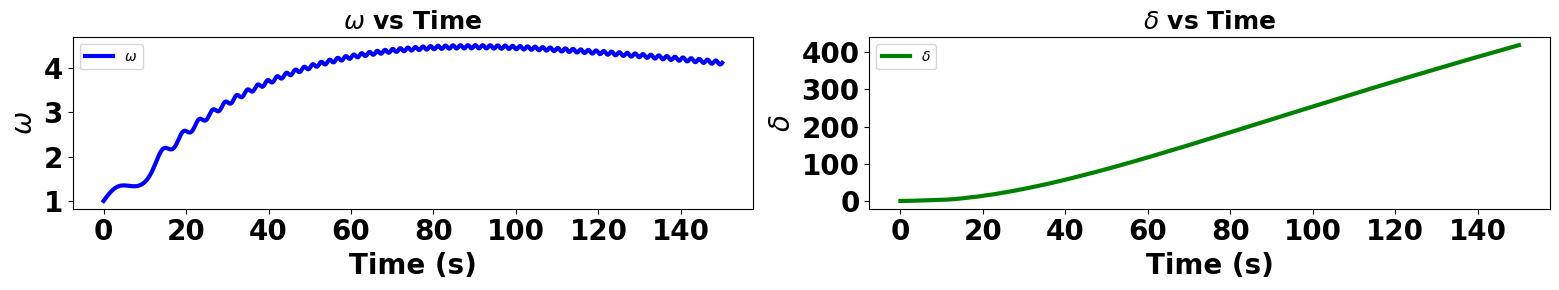}}
  \caption{Changing state variables $\omega$ and $\delta$ in sunrise to sunset solar variations when solar power increased 20\% than a normal sunny day.}
  \label{fig: fig3}
\end{figure}

These two equations describe the behavior of a power system with $n$ buses and $m$ generators. Equation (\ref{eq: 22}) represents the rate of change of the phase angle $\delta_i$ of generator $i$ relative to the system reference angle, where $\omega_B$ is the system base frequency, $\omega_i$ is the angular velocity of generator $i$, and $\omega_s$ is the synchronous speed of the generator. Equation (\ref{eq: 23}) represents the rate of change of the angular velocity $\omega_i$ of generator $i$, which is influenced by the mechanical power input $P_{mi}$, the electrical power output $P_{ei}$,  the solar power $P^{s}_{r}$, and the damping coefficient $D_i$ and inertia constant $H_i$ of the generator.

By incorporating the solar power term into equation (\ref{eq: 23}) and configuring the system parameters, we ensure the stability of the system. Now, let's delve into the two cases under discussion:

\textbf{Case 1: Sunny Day.} On a sunny day, the fluctuations in solar power throughout the day are negligible, tending towards zero. Under these conditions, we posit that the mechanical power ($P^{m}_{k}$) operates at 50\% efficiency. To counterbalance the electrical power, solar power ($P^{s}_{r}$) is incorporated, ensuring system stability. By increasing the value of \( P^{s}_{r} \) by 20\%, the system variables, as depicted in Fig \ref{fig: fig3}, become unstable with both \( \omega \) and \( \delta \) rising sharply. Given these conditions, the inherent stochasticity of solar power can be overlooked due to minimal fluctuations, allowing us to model the dynamic power system as an ordinary differential equation.

\begin{figure}[ht]
\centering
  \centerline{\includegraphics[width=0.5\textwidth]{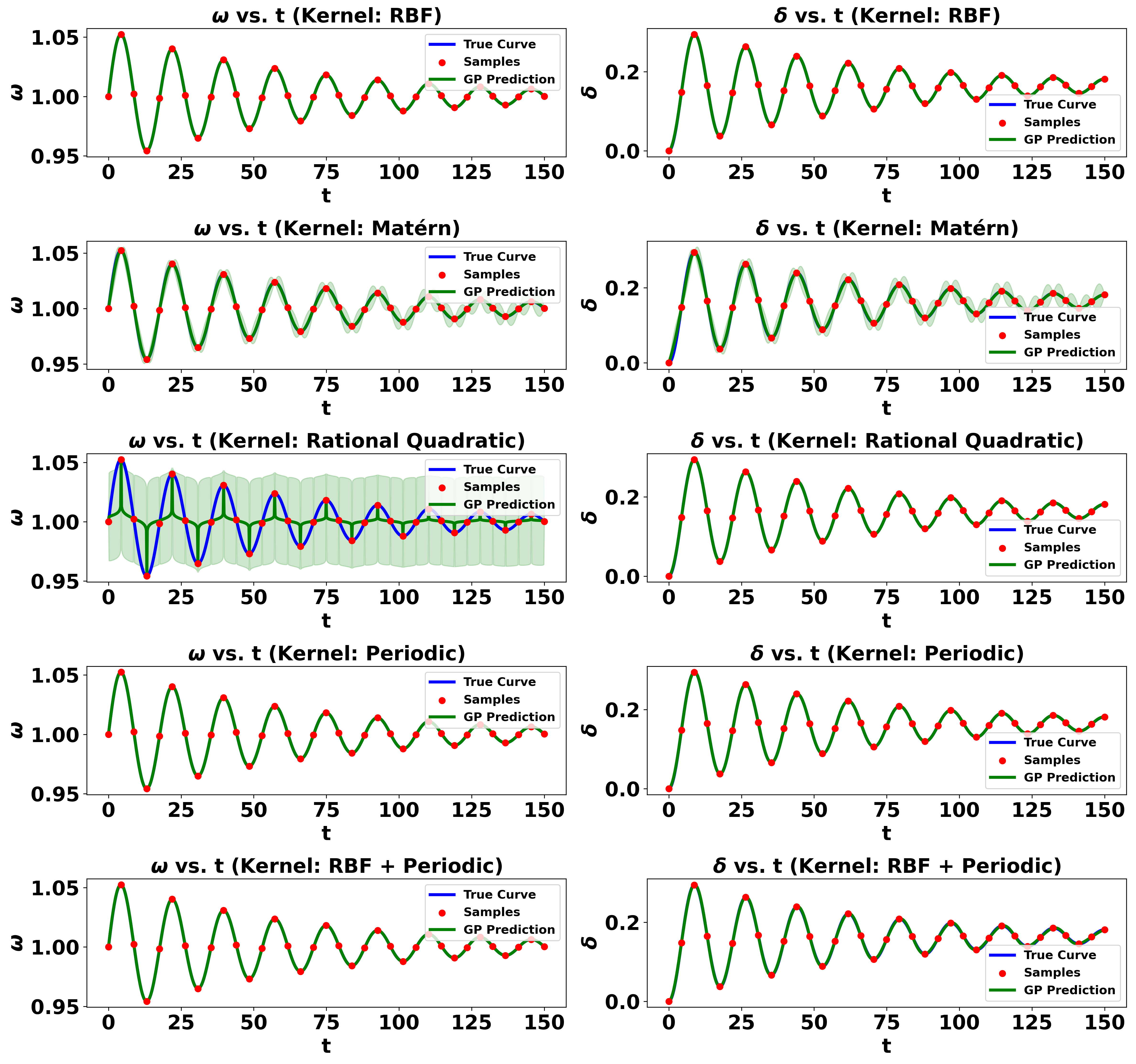}}
  \caption{Predicted state trajectory influenced by solar power on sunny days using various kernels.}
  \label{fig: fig4}
\end{figure}

Referring back to Fig \ref{fig: fig4}, as previously stated, we employed a mix of classic and ensemble kernel methods to assess the state estimation of the system under stable conditions. Notably, the ``RBF" and the ``ensemble model" yielded estimations closely mirroring the true function. In table \ref{tab:1}, the determination coefficient \( R^2 \) approaches 1, and both the Mean Squared Error (MSE) and Mean Absolute Error (MAE) for these kernels, relative to others, tend towards zero, indicating a commendable estimation performance.

\begin{figure}[ht]
\centering
  \centerline{\includegraphics[width=0.50
  \textwidth]{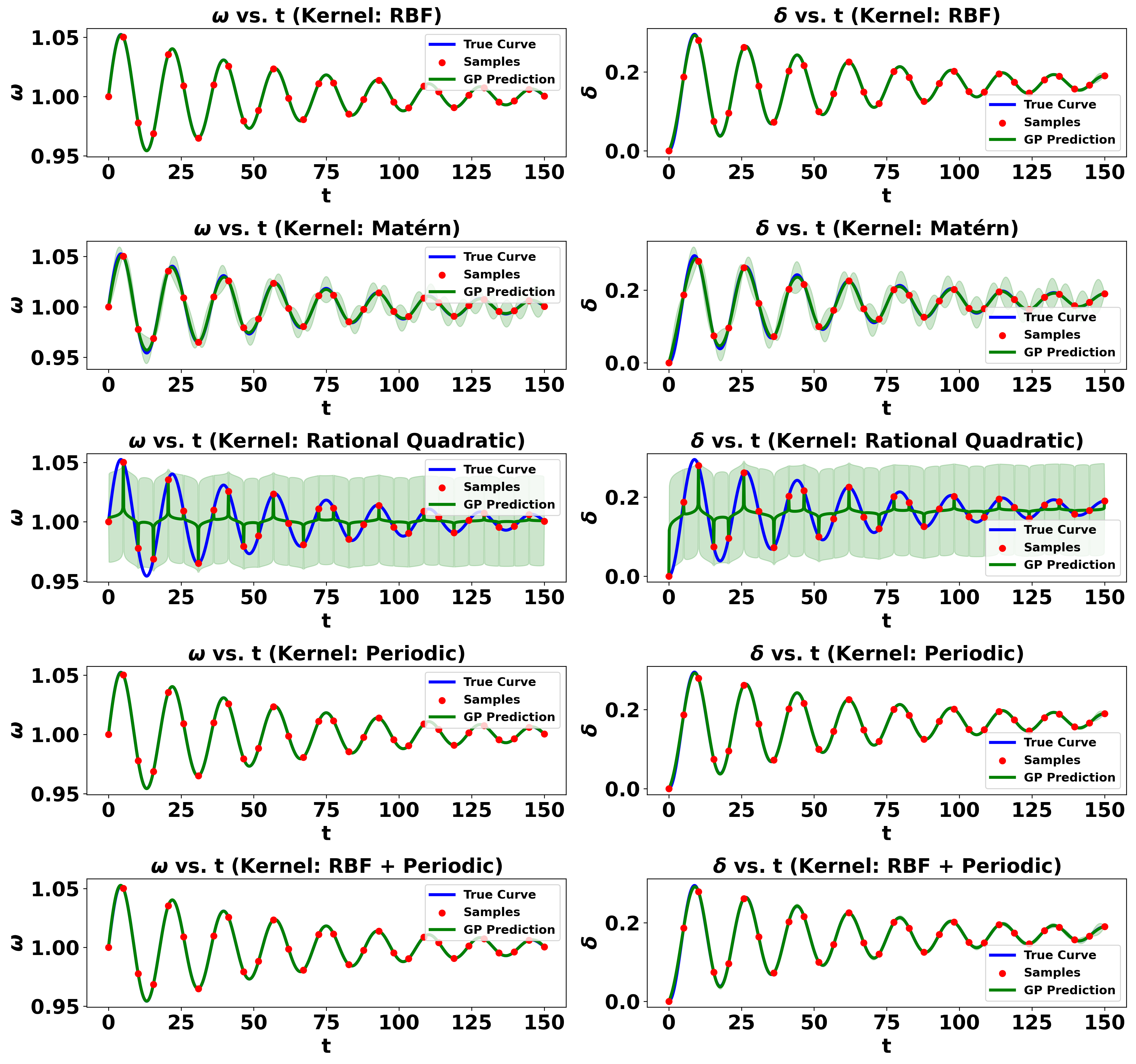}}
  \caption{Predicted state trajectory influenced by solar power on cloudy days using various kernels.}
  \label{fig: fig5}
\end{figure}

\textbf{Case 2: Cloudy day.} As depicted in Fig \ref{fig: fig1}, on a cloudy day, the presence of clouds intermittently obscuring the sun results in fluctuations in solar power throughout the day. Consequently, when integrating \( P^{s}_{r} \) into equation (\ref{eq: 23}), this term can be treated as stochastic, transforming the dynamic power system from an ODE to a stochastic dynamic equation. Contrasted with Case 1 under analogous conditions, the system exhibits heightened sensitivity to \( P^{s}_{r} \). With a 50\% efficiency of \( P^{m}_{k} \) and considering a range of \( (0.9-1.1) \times P^{s}_{r} \), system stability is attained, as demonstrated in Fig \ref{fig: fig5}. In this scenario, we also performed state estimation using the previously introduced kernels. The ``ensemble kernel" exhibited superior estimation accuracy in relation to the actual function. Referring to Table~\ref{tab:2}, all performance metrics indicate that this kernel provides a more precise estimation.
\vspace{-5pt}

\hspace{-6mm}

\vspace{-5pt}





\begin{table}[h]
\centering
\caption{Performance metrics for smooth solar power}
\label{tab:1}
\resizebox{\columnwidth}{!}{
\begin{tabular}{|c|c|c|c|c|c|c|}
\hline
\multirow{2}{*}{\textbf{Kernel}} & \multicolumn{2}{c|}{\textbf{MSE}} & \multicolumn{2}{c|}{\textbf{MAE}} & \multicolumn{2}{c|}{\textbf{$R^2$}} \\
\cline{2-7}
& \textbf{w} & \textbf{$\delta$} & \textbf{w} & \textbf{$\delta$} & \textbf{w} & \textbf{$\delta$} \\
\hline
RBF & 4.82e-07 & 8.92e-07 & 0.00016 & 0.00030 & 0.99996 & 0.999995 \\
Matérn  & 2.18e-05 & 7.30e-06 & 0.00108 & 0.00067 & 0.9984 & 0.99996 \\
Rational Quadratic & 3.49e-05 & 0.00119 & 0.00225 & 0.0147 & 0.9974 & 0.9935 \\
Periodic & 1.76e-05 & 7.30e-06 & 0.00099 & 0.00067 & 0.9987 & 0.99996 \\
RBF + Periodic & 3.00e-06 & 4.02e-06 & 0.00059 & 0.00051 & 0.99978 & 0.999978 \\
\hline
\end{tabular}
}
\end{table}

\begin{table}[h]
\centering
\caption{Performance metrics for noisy solar power}
\label{tab:2}
\resizebox{\columnwidth}{!}{
\begin{tabular}{|c|c|c|c|c|c|c|}
\hline
\multirow{2}{*}{\textbf{Kernel}} & \multicolumn{2}{c|}{\textbf{MSE}} & \multicolumn{2}{c|}{\textbf{MAE}} & \multicolumn{2}{c|}{\textbf{$R^2$}} \\
\cline{2-7}
& \textbf{w} & \textbf{$\delta$} & \textbf{w} & \textbf{$\delta$} & \textbf{w} & \textbf{$\delta$} \\
\hline
RBF & 3.24e-05 & 4.15e-05 & 0.00306 & 0.00369 & 0.9976 & 0.99977 \\
Matérn & 4.14e-06 & 4.19e-05 & 0.00102 & 0.00368 & 0.99969 & 0.99977 \\
Rational Quadratic & 0.00050 & 0.00147 & 0.01097 & 0.02226 & 0.9624 & 0.9920 \\
Periodic & 3.24e-05 & 4.19e-05 & 0.00306 & 0.00368 & 0.9976 & 0.99977 \\
RBF + Periodic & 3.92e-06 & 2.95e-05 & 0.00099 & 0.00233 & 0.99971 & 0.99984 \\
\hline
\end{tabular}
}
\end{table}

\vspace{-8pt}
\section{Conclusions}
We integrated solar power into the swing equation to supplement mechanical power in producing electrical energy. Two distinct scenarios for solar power were explored, and the results for the compact version of the 14-bus system, which consists of 5 generators, were presented as a single generator's output. Before solving the SDEs, we started with deterministic ODEs on sunny days to understand the baseline system behavior. We then added stochastic behavior of the cloud to better capture the influence of solar fluctuations. Various kernels were utilized for state estimation and short-term prediction of phase angle and angular velocity within an N-generator power grid system, relying on partial measurements.

\ifCLASSOPTIONcaptionsoff
  \newpage
\fi





\end{document}